\documentclass[runningheads]{llncs}
\usepackage[T1]{fontenc}
\usepackage{graphicx}
\usepackage{hyperref}

\begin{document}
\title{Generative AI \& Changing Work: Systematic Review of Practitioner-led Work Transformations through the Lens of Job Crafting}
\titlerunning{Gen AI \& Changing Work: Systematic Review}
% If the paper title is too long for the running head, you can set
% an abbreviated paper title here
%
\author{Matthew Law \inst{1}\orcidID{0000-0003-1167-9138} \and \\  Rama Adithya Varanasi  \inst{2}\orcidID{0000-0003-4485-6663} }
\authorrunning{Law \& Varanasi., 2025}
% First names are abbreviated in the running head.
% If there are more than two authors, 'et al.' is used.
%
\institute{Denison University, Granville OH, USA 
\and New York University, New York City NY, USA\\}
\maketitle              % typeset the header of the contribution
\begin{abstract}
Widespread integration of Generative AI tools is transforming white-collar work, reshaping how workers define their roles, manage their tasks, and collaborate with peers. This has created a need to develop an overarching understanding of common worker-driven patterns around these transformations. To fill this gap, we conducted a systematic literature review of 23 studies from the ACM Digital Library that focused on workers' lived-experiences and practices with GenAI. Our findings reveal that while many professionals have delegated routine tasks to GenAI to focus on core responsibilities, they have also taken on new forms of AI managerial labor to monitor and refine GenAI outputs. Additionally, practitioners have restructured collaborations, sometimes bypassing traditional peer and subordinate interactions in favor of GenAI assistance. These shifts have fragmented cohesive tasks into piecework creating tensions around role boundaries and professional identity. Our analysis suggests that current frameworks, like job crafting, need to evolve to address the complexities of GenAI-driven transformations.

\end{abstract}

\section{Introduction}
Like any large-scale change, technological innovations, such as the emergence of personal computers, smart devices, and the advent of the ``first wave''~\cite{Autor2023} of AI-integration, have contributed to destabilization in workers' lives \cite{Anteby_2016}. Destabilization is not necessarily negative, and technology-enabled destabilization can sometimes create positive outcomes for workers, such as increased productivity \cite{Brynjolfsson2023}. But for the majority of workers, these changes create a period of vulnerability and challenges that compel modification of their practices to keep themselves relevant. Studying these \textit{worker-led transformations} in response to the evolving techno-centric labor landscape can illuminate how practitioners might not only regain but maintain stability despite occupational destabilization. 

While blue-collar workers have been the most significant victims of the aforementioned occupational destabilization~\cite{Ford_2015},  GenAI (e.g., ChatGPT) is destabilizing white-collar occupations~\cite{Felten2023} as well.
%\cite{Kalleberg_2009} as well. 
%In fact, GenAI is poised to shape and impact as many as 80\% of occupations in the U.S. alone \cite{Eloundou_2023}. This scale of impact is unprecedented. 
Emerging studies on the effect of GenAI tools in this space tend to take a top-down approach, limiting workers' experiences to narrow metrics within controlled settings \cite{Brynjolfsson2023}. In contrast, recent HCI studies have taken a bottom-up approach to capture more nuanced worker experiences. While these studies offer valuable insights, they are typically limited to specific occupations, highlighting the need for a broader understanding across various fields. Addressing this gap, our research objective was to \textbf{understand how practitioners across diverse occupations have transformed their own work in response to GenAI integration}. 

We conducted a systematic literature review of 23 papers in the ACM Digital Library from the past two years (2022–2024). This period marked the widespread adoption of off-the-shelf GenAI tools. Through this review, we aimed to develop generalizable insights into the lived experiences of practitioners.
%, which could help workers safeguard themselves from potential harms of GenAI while enhancing their overall job outcomes.
Our findings show that, with the integration of GenAI into their work, practitioners across 18 professions transformed their \textit{tasks} in various ways. Many delegated peripheral tasks to GenAI, freeing up time for higher-priority core activities, while some actively engaged with GenAI to enhance the quality of core tasks. In certain instances, practitioners even assigned core tasks to GenAI. However, this flexibility brought with it a new burden: the need for ``AI managerial labor'': additional work to oversee and manage GenAI effectively.

Practitioners also transformed their \textit{collaborations} by increasing their reliance on GenAI where they might have otherwise relied on other stakeholders. In extreme cases, practitioners used GenAI to bypass peers and subordinates to circumvent negative outcomes or achieve positive ones. 
%These kinds of GenAI-driven reconfigurations could foster grappling with changing roles and identities. For example, practitioners could view themselves as assistants to GenAI when offloading core tasks or managers when they delegating peripheral tasks. 

Our findings highlight the emergence of several tensions. First, while practitioners adopted GenAI to reduce their workload, resulting demands to manage GenAI and its outputs shifted their roles away from their professional identities. Second, applying GenAI to specific task components fragmented traditional workflows into piecework, eroding established boundaries and safeguards, with certain tasks being absorbed into other roles. Our study makes the following contributions:
\begin{enumerate}
    \item We present comprehensive, worker-led transformations in identity, tasks, and relationships across 18 professions in response to the introduction of GenAI.
    \item We group and characterize various forms of emerging GenAI-specific work undertaken by practitioners under the umbrella of \textit{AI managerial labor}. 
    \item We lay out tensions that emerged as a direct result of practitioners integrating GenAI into their workflows. 
\end{enumerate}

\section{Related Work}

\subsection{AI's Role in Work Transformations}

While the relationship between automation and work has long been studied,
Frank et al.\cite{frank2019toward} argue that the impact of AI-driven trends on labor may be harder to predict.
Autor~\cite{autor2015there} notes that automation tends to increase the value of complementary tasks where humans have comparative advantages.
% Autor~\cite{autor2015there} has argued that automation tends to complement tasks that it cannot substitute for, potentially leading to a bifurcation whereby automation hollows out middle tasks and creates demand for high-level abstract tasks and manual labor. However, Autor notes that machine-learning-driven AI, using inductive methods, diverges from traditional automation techniques in ways that might threaten tasks that utilize tacit knowledge, which traditionally have resisted automation. This raises the question of how and to what extent AI technologies might transform human work. 
%Through a quantitative lens, Park and Kim analyzed automation at a task level from 2008 to 2020, finding that machine control and hazardous tasks have become more automated, while systems analysis and creative and critical thinking have become less so \cite{park2022data}. However, a fair amount of work has also tried to answer this question through the experiences of workers in domains that have begun to integrate AI.
Early work in HCI suggests that, indeed, AI's limitations favor complementary transformations of work rather than simply displacing human labor. Fox et al. describe patchwork as a form of new invisible labor that emerges around the gaps between AI expectations and reality~\cite{fox2023patchwork}. This labor takes different forms, including \textit{compensating patchwork}, where humans fill in for AI when it breaks down, \textit{peripheral patchwork}, where humans watch over supposedly autonomous machines, and \textit{collaborative patchwork}, where a human works in concert with the AI to help adapt to contextual changes like the weather. However, how these changes will affect the dynamics of work across different professions remains to be seen.

%Workers also create patches--manual reconfigurations of AI systems to make them more reliable.
%In a similar vein, Levy observes that, while future models of automation in trucking may delegate more tasks completely to AI, the current reality is one of integrating AI into truckers' work through \textit{compelled hybridization} like smart wearable or camera systems to monitor fatigue while driving~\cite{levy2023}.
% Outside the realm of manual work, Wang et al. spoke to data scientists about the advent of AutoAI, which is geared towards automating data science tasks. Overall, they found that the data scientists viewed continued automation in their field as inevitable, but also believed that human expertise would never be fully displaced~\cite{wang2019human}.

% Overall, the evidence seems to suggest a near-term a divide between the vision of AI-driven automation in the workplace and the messy reality of how integrating AI into human work actually changes that work. With that said, the nature and availability of AI technologies in the workplace have drastically changed over the past two years with the rise of generative AI models. 

% \subsection{GenAI's Early Signs of Impact on Working Communities}

Rapid innovation in GenAI and their multi-modal task capabilities positions them to support professional work in new ways.
%such as hospitality \cite{Hsu_Tan_Stantic_2024},education \cite{Chan_Hu_2023}, and manufacturing \cite{Bendoly2023}. 
%\cite{23Jiang,Small2024}. 
Recent research predicts that almost 80\% of the U.S. workforce could have at least 10\% of their work tasks altered by GenAI technologies~\cite{Eloundou_2023}, exposing even high-skill roles to GenAI~\cite{Felten2023}.    

Two research narratives have emerged. The first has explored the potential of GenAI to bolster workers' tasks, skills, and outcomes. Experimental studies have indicated positive worker performance \cite{Al2024,Brynjolfsson2023}. 
%For instance, an initial study with customer support agents using Gen AI reflected a 14\% increase in the number of issues that they resolved per hour\cite{Brynjolfsson2023}. 
% A more recent study exploring the impact of GenAI tools on professional work in ten different job categories demonstrated at least 26 \% savings in their task duration\cite{Dell2023}. Here, professional consultants in a controlled experiment either delegated portions of work or continually interacted with GenAI, indicating nuances in their GenAI usage. GenAI tools might also augment employees' creative processes by provoking idea generation through serendipity~\cite{Epstein_2022}. 
Another stream of research has investigated shortcomings of GenAI technologies that negatively impact aspects of work. Recent studies have shown that professionals using GenAI may introduce biases and errors into their work processes, affecting the artifacts they produce %\cite{Miyazaki2024,
\cite{Kidd_Birhane_2023}. More broadly, studies have argued the negative impacts of these micro-changes on the development of professional capacities, with a strong potential to dilute human skills, such as creativity and critical thinking \cite{Walczak_2023}. 

\subsection{HCI, Work, and Gen AI}
Several projects have already examined how specific communities or fields are engaging with GenAI. One question of interest is how practitioners in different fields are adopting such tools.
For example, Takaffoli et al., studying UX designers, found, among other things, that they used GenAI more for research tasks than design tasks, and more at an individual level than at a team level~\cite{takaffoli2024generative}.

Some projects focused on how GenAI becomes adopted in a given context. Boulus et al.~\cite{boulus2024genai} identified five stages in developing a functional augmentative relationship with GenAI.
%(1) playing around, (2) infatuation, (3) commitment, characterized by more serious considerations of how to use the tool, including financial and ethical considerations, (4) frustration with the tool as continued use exposes limitations or friction due to changes in the human or the AI's behavior, and (5) enlightenment, characterized by realization and adjustment to the complementary imperfections of themselves and the tool.
Boucher et al. \cite{22}, meanwhile, found initial resistance among game design interns who were encouraged to use GenAI by program directors. Interestingly, they found that interns were more likely to embrace its use in programming tasks than artistic ones, considering ethics and quality.
%Their reasoning focused on ethical questions like ownership and agency, as well as variance between the quality of work produced by GenAI across the two domains. 
% The authors also found differences in perception of GenAI as a tool, rather than a threat to replace them~\cite{22}.

Others examined practitioner perceptions of how GenAI will transform their industries.
Li et al. found that experienced UX designers were confident that their uniquely human abilities would relegate GenAI tools to assistive roles, but worried that increased usage of these tools could negatively impact junior designers~\cite{2}. In particular, they worried that pressure to keep up with the speed of GenAI  could lead to ``creativity exhaustion'' among human designers. 
%Kalving et al., through surveys, focus-groups, and interviews with designers, found a generally optimistic view of AI as a collaborator for human designers, couched in ethical concerns over issues like authenticity, and marked by an acceptance that designers would need to adapt, perhaps moving away from traditional UI design and focusing on uniquely human capacities like empathy for users~\cite{kalving2024ai}.

We were primarily interested in how practitioners are using GenAI on their own to transform the nature of their work and experiences. In particular, we were interested in the following research question (RQ): \textbf{what are the practitioner-led work transformations in response to the introduction of GenAI in their jobs?}

\subsection{Job Crafting: A Theoretical Lens} \label{RW-jobcrafting}
To address this question, we applied the theoretical framework of \textit{job crafting}, which originated as a theory to understand the design of people's work responsibilities, activities, and relationships. 
%What sets job crafting apart from other work design frameworks is that it attributes agency to workers, rather than managers or the organization, in designing their work. The framework focuses on the bottom-up physical and cognitive changes that workers initiate to modify their role or relationships boundaries in at least a semi-permanent way \cite{Tims_2012}. 
The idea was introduced to describe how professionals invest significant energy in configuring their jobs \cite{Wrzesniewski_2001}. A set of studies identified three main aspects of job crafting. First are the \textit{physical task} changes to the job that affect the `form' or the `number of activities.' Second are \textit{cognitive} changes that affect how individuals see the job. Third are \textit{relational} changes that influence interactions with colleagues. 

% Another strand of job crafting research views jobs as sets of demands (e.g., a tight deadline imposed on an journalist) and resources (e.g., a journalist using a fact-checking tool to make their research easier) and describes how in job crafting, workers attempt to reduce their demands and increase their resources.
Zhang et al.\cite{zhang_2019} synthesized job crafting research by suggesting two higher-order constructs: (1) \textit{approach} crafting, which involves actively creating opportunities that align with one's professional preferences, strengths, and goals, and (2) \textit{avoidance} crafting, which describes practices employed to reduce negative work outcomes. 

Workers' crafting behaviors can occur in response to either proactive (e.g., an author wanting to publish a new experimental work) or reactive motivators (e.g., a writer coping with a new change in the organization) \cite{Lazazzara_2020}. 
% Interestingly, workers who engage in crafting behaviors proactively have been shown to benefit with respect to increased job performance, satisfaction, and wellbeing \cite{Petrou2012,Gordon2018}. 
% Additionally, not all workers in a job setting engage in job crafting at the same level. This is quite diverse based on the rank \cite{Berg2010}, degree of autonomy \cite{Berg2010}, and the resources the workers \cite{Petrou2012} have available to them. These combined insights indicate that job crafting is a complex process that is unique to every worker depending on their job \cite{Petrou2012}.
%their role, and their work environment, which can become a part of their daily work lives \cite{Petrou2012}, especially to improve meaning and increase moments of hopefulness \cite{Blustein2023}.
Job crafting theory has been studied extensively with full-time employees in traditional organizational contexts across various domains \cite{Berg2010,Fuller2017}. 
% Within these settings, studies have focused on both proactive and reactive crafting practices in specific work situations, such as  shaping identity \cite{Kenny2020}, goal-setting \cite{Bruning2019}, planning and scheduling \cite{Kossek2016}, work-transitions \cite{Gascoigne2018}, work-life balance \cite{Gravador2018}, and developing professional relationships \cite{Bizzi_2017}.
With recent transformations in job structures, job crafting has also been studied in contexts of non-traditional forms of work, such as gig work \cite{Wong_Fieseler_2021}. 
% For example, \cite{Panteli2022} explored how female IT contractors craft their jobs to thrive in a male-dominated sector. Similarly, \cite{Wong_Fieseler_2021} found that gig workers who engaged in individual and collaborative job crafting behaviors indicated higher resilience. 

Despite the burgeoning interest in job crafting, there is limited understanding of how AI technologies are shaping workers' crafting perspectives. Studying changing worker-led practices through the lens of job crafting in the advent of GenAI technologies is beneficial for multiple reasons. The job crafting lens allows us to position agency in the hands of the workers, the actual users of the GenAI technology. This framework also allows us to balance some of the techno-deterministic narrative surrounding the GenAI push with a social constructivist understanding of how individuals integrate GenAI into their work. Lastly, it provides a systematic approach to examine the nuanced sociotechnical practices at a micro-level.

\section{Methods}
We performed a systematic literature review using the thematic synthesis approach to answer our broad research objectives \cite{Thomas2008}. Rather than beginning with a presupposed research question or hypothesis, thematic synthesis enables researchers to start with a high-level objective and use a systematic, inductive analysis to provide comprehensive answers while iteratively refining the objective. This method is particularly suited for reviews aimed not only at summarizing but also at \textit{extending} the existing literature through synthesis, producing higher-order structures between broad concepts \cite{Xiao_Watson_2019}.
% In this sense, the study draws on the long-standing tradition of meta-synthesis, such as meta-ethnography that focuses on producing inferences that expand the current understanding of the literature \cite{Thorne2004}.

% \begin{figure}
%     \centering
%     \includegraphics[width=0.5\linewidth]{images/cscw_analysis.png}
% \caption{A flowchart showing the overall steps taken to screen and shortlist the key literature for analysis}
%     \label{fig:analysis-process}
% \end{figure}

Thematic synthesis is conducted in three steps \cite{Thomas2008}. First, researchers identify, collect, and filter relevant literature to build a study corpus. Second, the study corpus is analyzed to extract core descriptive themes through the coding process, also known as second-order constructs. Third, the themes are clustered and synthesized into analytical themes, or third-order constructs.

\subsection{Searching \& Scoping the Literature}

\subsubsection{Identification}
We began the research process with a broad objective: understanding the relationship between recent AI developments and worker practices. Before initiating the main search, we conducted a preliminary exploration of the research landscape. We looked for papers relevant to AI and work practices in four prominent digital repositories containing HCI work: ACM, Elsevier, Springer, and IEEE. In parallel, we reached out to 15 expert researchers in the fields of technology, organizational studies, and labor studies, asking them to recommend relevant papers and key research gaps. We identified these experts from specific academic groups, mailing lists, prior workshops, and online communities. We shared a brief excerpt of our research objective and our proposed approach before asking them to suggest two to four literature items, excluding gray literature items such as workshops and work-in-progress.

Our preliminary exploration and expert recommendations revealed that papers describing how workers engaged with AI tools and adopted the technology in a bottom-up manner seemed the most relevant to our objective. We also observed a lack of consensus among the qualitative studies regarding how AI use was shaping practices beyond particular working groups. In light of these insights, we refined our final search to reduce the scope exclusively to how workers employed GenAI tools and the bottom-up, proactive efforts through which they used these to transform their work identity, practices, and relationships, for which the ACM Digital Library yielded the most relevant studies.

\subsubsection{Collection}
Based on our refined objective, we started our focused search in the ACM Digital Library through two main areas of keyword search: ones that described and captured \textit{GenAI} technologies and ones that captured practitioner's bottom-up \textit{work practices}. We experimented with the keywords until we achieved a good balance of finding diverse papers while minimizing noise, motivated by a purposive sampling strategy ~\cite{Xiao_Watson_2019}. Purposive sampling focuses on studies that can assist in conceptual synthesis through rich \textit{interpretation} rather than exhaustive search. Table \ref{tab:search_query} presents the final keywords used to find the initial corpus. We constructed our query as the conjunction of the disjunction of terms in each of these areas, over both the title and abstract fields. %Thus, an entry containing at least one of the terms in each area, in either the title or the abstract, was contained in our result set. 

\begin{table}
    \centering
    \begin{tabular}{p{1cm} p{1 cm} p{7.5cm} p{1.3cm}}
        \hline
       \textbf{KC} & \textbf{Op}  & \textbf{Search Terms}   & \textbf{Scope}\\
        \hline
        GenAI &
        &  
       ``generative artificial intelligence'' OR ``generative ai'' OR ``gen AI'' OR ``genAI'' OR ``large language models'' OR ``chatgpt'' OR ``chat-gpt'' OR ``stable diffusion'' OR ``dall-e'' OR ``midjourney'' OR ``AI-generated content'' OR ``text generation AI'' OR ``image generation AI'' OR ``AI creativity tools'' OR ``AI art generation'' OR ``AI-enhanced tools''&  
      
        Title,\ \ \ \ \ \ \ \ \ \ \ \ Abstract\\
       &  (AND) & & \\
        Work Practices &
        &
         ``labor'' OR ``jobs'' OR ``worker'' OR ``practitioner'' OR ``professional'' OR ``staff'' OR ``workforce'' OR ``work practices'' OR ``employment'' OR ``occupational'' OR career OR ``work habits'' OR ``job performance'' OR ``work design'' OR employee OR ``work patterns'' OR ``work routines'' OR ``work strategies'' OR ``employment practices'' OR ``task management'' OR ``job adaptation'' OR ``work changes'' &
         Title,\ \ \ \ \ \ \ \ \ \ \ \ Abstract\\     
        \hline
    \end{tabular}
    \caption{Final search query used to find relevant papers in the ACM Digital Library. The key concepts (KC) were joined by an AND operation (Op) and search terms were separated by OR operation. These keywords were applied to titles and abstracts.}
    \label{tab:search_query}
    \vspace{-.42 in}
\end{table}

\subsubsection{Screening}
Our search query included results from 2022 to 2024 (until July), yielding 665 results. From an initial dataset of 665 papers, we performed a pass to identify the most relevant papers in two stages: 1) title screening and abstract screening, and 2) full article screening. To screen the papers effectively, we followed clear inclusion and exclusion criteria. We focused on only those experiences that described and presented evidence of change in work practices, avoiding speculative work or initial experiences from the system pilots. Papers were included if they talked about a specific working community, their work practices, and the use of GenAI in these practices. We present a more detailed list of inclusion and exclusion criteria in Table \ref{tab:exclusion_inclusion}.

While we mainly focused on qualitative studies over papers focused on lived experiences, we included some survey and/or mixed-method papers, including those that also designed interventions around what they learned.
% mixed-methods papers that used surveys or described intervention designs around experiences. 
Thematic analysis is particularly well-suited to analyzing and synthesizing such rich qualitative research, enabling both consensus-building and the development of extended arguments \cite{Barnett2009}. 
  
The title and abstract screening process was divided between two reviewers who rated papers independently. To reduce preconceived biases, we did not include authors' names or affiliations. To kick-start the process, both authors reviewed a smaller sample of papers together to build consensus on how to apply the filtering criteria. Once the authors started independent review, they used prolonged deliberation \cite{creswell} as a method to discuss, mark, and resolve uncertain cases. During the title and abstract filtering stage, we reduced the set of 665 papers down to 27. During our subsequent full paper review stage, we eliminated four more papers using the same criteria, resulting in a final set of 23 papers.

\begin{table}
    \centering
    \begin{tabular}{p{5cm} p{7cm}}
        \hline
        Inclusion Criteria   & Exclusion Criteria \\
        \hline
        \begin{enumerate}
            \item Studies that captured workers' change of practices as one of their key themes.
            \item Studies that focused on GenAI.

        \end{enumerate} &
        \begin{enumerate}
            \item Studies that focused on speculative experiences.
            \item Papers that purely focused on computational methodologies. 
            \item Papers that covered experimental studies (e.g., comparative studies between practitioners' use and non-use of GenAI).
            \item Papers that focused on GenAI-based system design and its evaluation through practitioner feedback.
            \item Papers that conducted systematic review.
            \item Low-quality studies.    
            \
        \end{enumerate}
        \\
        \hline
    \end{tabular}
    \caption{Table depicting the inclusion and exclusion criteria for finding relevant papers.}
    \label{tab:exclusion_inclusion}
    \vspace{-.2 in}
\end{table}

\subsection{Analysis}
We started our analysis by familiarizing ourselves with the kinds of reported data in the papers and their style of reporting. 
%All the papers in our final list followed a consistent structure of reporting, with most of their empirical data being available in the findings, in the form of key concepts and quotations. This aligned with the recommendations suggested by \cite{Thomas2008}'s best practices. 
Afterward, the authors went through each paper's finding section line-by-line, engaging with their concepts and coding the key insights in order to develop second-order constructs. Once we developed codes for a few papers, we used peer debriefing \cite{creswell} to discuss the codes and work on the disagreements. This process was repeated until all the papers were coded. Throughout this and the following process the authors repeatedly discussed, refactored, and updated the codes, ultimately arriving at a final set of 24.

In the second phase of our analysis, we focused on developing third-order constructs that went beyond the data presented in the papers. We used an abductive approach, utilizing the theoretical framing of job crafting \cite{Wrzesniewski_2001} to produce interpretations that mapped to our research objective. For instance, the we employed the notion of task and cognitive crafting to analyze how practitioners reconfigured their tasks and repositioned their beliefs around job identity. This process was repeated until comprehensive analytical themes emerged. To help with the process, we developed multiple conceptual maps that helped us draw connections across the descriptive themes \cite{Miles2020}. This analysis ended with \textit{thirteen} analytical themes, including \textit{`AI managerial labor,'} \textit{`Expanding role capacities,'} and \textit{`Displacing human dependencies with GenAI'.}\footnote{The final codebook is available at \href{https://osf.io/nprc9/}{https://osf.io/nprc9/}}  
%A full accounting of the key themes and relevant codes are presented in the attached supplementary materials. 

\section{Findings}
To organize our findings, we start by describing high-level findings, including the key characteristics of the practitioners. Next, we present how practitioners used GenAI technologies to transform their overall work, specifically their work processes, role attributes, and professional relationships. 
%Using the theory of job crafting \cite{Wrzesniewski_2001,tims2010job} as an anchor, we start by examining how individuals shaped their tasks and responsibilities through task crafting techniques. We then examine how practitioners have molded their professional networks through relationship crafting techniques. Lastly, we show how practitioners use cognitive crafting techniques to make meaning of their changing roles.
\begin{table}
    \centering
    \begin{tabular}{|c|p{0.7\textwidth}|}
        \hline
        Occupations & Software developer, manager, lecturer, data engineer, UX designer, UX researcher, UX writer, industrial design,  artists,  UI designer, social worker,  architect, fact-checker, knowledge worker, fiction writer, researcher, speech-language pathologist, executive \\
        \hline
        Industries & Education, technology, art/culture, design firm, agroindustry, health,  IT, fact checking, gaming, science fiction, research \\ 
        \hline    
    \end{tabular}
    \caption{Paper Demographics}
    \label{tab:paper_demographics}
    \vspace{-.3 in}
\end{table}

% The papers included in our analysis covered 18 different occupations across 11 different industries. The majority (n=15) focused on practitioners in technology-focused or design roles, although a few papers (n=8) also studied roles outside this archetype, including artists, speech-language pathologists, and writers. Sometimes these lines blurred, e.g., in game design settings involving developers and artists. 

The papers included in our analysis covered 18 different occupations across 11 different industries. The majority focused primarily on practitioners in technology or design-focused roles (n=15). However, we found a few papers that also investigated one or more roles outside this archetype, including artists, speech-language pathologists, and writers (n=8). The sectors in which these practitioners worked spanned from technology and design to fact-checking, research, and health. For a full breakdown of the occupations and industries included in our survey, please see Table~\ref{tab:paper_demographics}.

In the papers examined, practitioners used a wide range of GenAI tools, including off-the-shelf products such as ChatGPT, Midjourney, Copilot, and LLaMA. The modalities of these tools varied, spanning text (e.g., an LLM-driven chatbot), audio (e.g., voice), images (e.g., a text-to-image generator), and code (e.g., an LLM-driven coding assistant). While many practitioners used standalone tools like Midjourney, others utilized GenAI-enabled features integrated into commercial platforms, such as Notion or Figma plugins~\cite{6}, or custom-developed tools designed for specific purposes, such as a chatbot that interacts with patients to populate a health dashboard~\cite{5}.

In the following subsections, we describe different transformations practitioners brought in their work to incorporate GenAI\footnote{A detailed table of these transformations is available at \href{https://osf.io/nprc9/}{https://osf.io/nprc9/}}. Throughout our findings, we differentiate between avoidance crafting and approach crafting. Avoidance crafting refers to practices undertaken by an individual to reduce negative work outcomes, while approach crafting refers to the active creation of opportunities that align with one's professional preferences, strengths, and goals~\cite{zhang_2019}.

\subsection{Transforming Work Processes}
\label{sec:workprocesses}
Practitioners primarily transformed their work by leveraging GenAI to craft and reshape various stages of their regular tasks. This concept is known as task crafting, and our analysis revealed it was utilized across seven different stages of work: discovery, analysis, artifact creation, quality assurance, delivery, logistics, and overall management of work.
%(see Figure \fixme{\ref{fig:task-processes}}).

% The second category focused on developing the practitioner's own professional role, both in short- and long-term. The third category focused on the crafting labor that went into the technology to fine-tune and customize it to improve its efficacy the crafting process. 

\subsubsection{Discovery}
A significant proportion of the transformed practices observed in the papers we reviewed involved tasks in the discovery phase of the participants' work. The discovery phase consists of early-stage processes where practitioners learn more about the problem at hand. One way practitioners used GenAI in the discovery phase was by delegating relevant efforts to it, reducing their scope (avoidance crafting). A common theme in this category was offloading time-consuming information foraging tasks to GenAI. For example, a fact checker in~\cite{1} described how they delegated their laborious work of finding trending topics to ChatGPT:

\begin{quote}
    \textit{``We take the top 200 headlines from the last 24 hours from those sites [...] and run them through ChatGPT, asking it to summarize the main narratives [...] and extract the names of people, places, entities [...] and then send that to me by email. So every six hours, [...] we get an email.''}
\end{quote}

Software developers applied similar strategies to more specific instances of information foraging, such as aggregating syntax examples for guidance~\cite{4}. 
%All of these examples reduced the effort needed from practitioners to complete their tasks.

In contrast to delegating effort, several practitioners actively engaged with GenAI tools to improve aspects of their tasks in the discovery stage (approach crafting). This included practitioners seeking inspiration from GenAI as they embarked on projects. For example, developers used GenAI to seek out new problems to solve~\cite{4}, while visual artists engaged with GenAI playfully to find serendipitous inspiration~\cite{9}. It should be noted that seeking inspiration through GenAI was not limited to initial exploration. A participant in~\cite{6} used GenAI to help them flesh out ideas that they had already partially formed, while designers in~\cite{6,18} used GenAI to help develop mood boards. Developers, on the other hand, used it to explore new ways to solve problems, including learning complex logic creation and identifying ways to reuse solutions \cite{4}.

Another common activity observed in this category was using GenAI for brainstorming, with participants finding ways to speed up idea generation~\cite{10}. This involved coaxing responses or even including GenAI in their existing brainstorming techniques like reverse-thinking~\cite{21}. For example, a user interface designer \cite{6} explained using used GenAI to expand the scope of their ideation:   

\begin{quote}
    ``\textit{I would ask him to tell me alternatives. Push it to think about it in a different way. I always ask – any other ideas? And it would always come back with something}.''
\end{quote}

Practitioners also used GenAI to kick-start their creative processes, such as as generating a first draft of a written project ~\cite{10} or providing guidance on what kinds of software libraries might be useful for a problem~\cite{12}. 

%As one UX designer ~\cite{2} put it: 

% \begin{quote}
%     ``\textit{GenAI tools can help nowadays in generating basic ideas to help us populate the blank canvas, thereby aiding in overcoming the fear of the `empty canvas.'}''.
% \end{quote}

Additionally, GenAI helped kick-start work in areas where practitioners' confidence in their expertise wavered, such as an amateur programmer learning programming concepts to help them write code for a new project~\cite{6}, or software developers turning to GenAI to develop deeper understandings of their craft and to strategize solving challenging problems~\cite{16}.

\subsubsection{Analysis} Certain practitioners leveraged GenAI to change how they performed analysis-focused tasks. These tasks typically involved taking raw information and finding specific patterns. Similar to the discovery phase, practitioners delegated analysis tasks to GenAI that they felt were particularly time-consuming. For instance, a UX researcher~\cite{10} used a GenAI tool embedded in Miro\footnote{www.miro.com},
a digital whiteboarding tool, to summarize brainstorming data.

Practitioners also \textit{actively} engaged with GenAI to make their analysis process more efficient. For instance, \cite{10} showed how end-user-facing practitioners repeatedly consulted with GenAI to synthesize insights from different forms of raw data. A product designer shared how they did the analysis:

\begin{quote}
    \textit{``It was like 75 open ended survey responses and I [...] strip them of [the participant number] and dump them into ChatGPT, and asked [ChatGPT] to generate 5 insights based on the 75.'' }
\end{quote}

The UX practitioner in~\cite{21} engaged with GenAI for a variety of text analysis activities as well, from thematic analysis to identifying pain points in customer emails.

\subsubsection{Artifact Creation} 
In the intermediate to late stages of their processes, users relied on GenAI to help create work-related artifacts. The types of artifacts varied, including personas generated from user transcripts~\cite{21}, product descriptions~\cite{21}, merchandise, and game assets~\cite{20}. Surprisingly, the practices tended to lean toward avoidance crafting rather than active collaboration with GenAI. The level of delegation ranged greatly. On the conservative end of the spectrum, some practitioners outsourced only a small portion of artifact creation, e.g.: 

\begin{quote}
``\textit{I could generate any of this in Photoshop or Illustrator [...] the fact that it was able to render these things on an aesthetic level that was exceeding my bar or at my bar, and doing it in an instant, was mind-boggling. What it did is it gave me time to dabble in other areas.}''~\cite{6}
\end{quote}

On the far end of the spectrum, practitioners used GenAI to perform the entire artifact creation, e.g., software developers asking GenAI to generate entire new features or classes from scratch~\cite{19,11}. Amplified output capacity was one of the most positive consequences of task crafting. A set of professionals in diverse roles at a game design firm, for example, ``\textit{were unanimous in considering the creation of more content in less time a strength of [GenAI] systems }''~\cite{9}. 
%~\cite{3} particularly associate this effect with content creation and creativity in ways that could influence practitioner's self-perceptions: ``by leveraging ChatGPT's capabilities to expand their creative or informational output, participants experienced a sense of productivity and accomplishment.''

\subsubsection{Quality Assurance \& Delivery} 
Beyond creating task-related artifacts, multiple papers provided evidence of participants using GenAI for quality assurance and delivery tasks. These practices were primarily aimed at offloading participants' responsibilities (avoidance crafting). For quality assurance tasks, this often meant relying on GenAI for code maintenance in software development, such as refactoring code snippets~\cite{4,19}. Those in user-facing roles, such as UX practitioners, focused on designing evaluation methods to maintain or even improve product quality. For example, some used GenAI to generate user test case scenarios, directly integrating the scenarios' output into their projects~\cite{21}.

Reliance on GenAI increased during the delivery stage, with practitioners delegating the creation of resources meant to communicate key ideas to stakeholders. These resources included pitch documents to present content strategy, images to visualize concepts, and presentations to aid communication during handover~\cite{9,10,12,21}. UX designers in~\cite{10,21} described using Midjourney to generate visual representations and mock-up descriptions, all with the goal of reducing communication challenges and enhancing clarity. Developers, in turn, leveraged GenAI extensively for creating various forms of code documentation~\cite{4,19} and reports~\cite{8}.
% This was particularly evident in~\cite{8}, where software developers used GenAI to minimize the effort required for writing reports.
% One developer shared:

% \begin{quote}
% \textit{``The tool’s ability to generate coherent and relevant content and provide valuable insights and suggestions significantly supported participants’ writing activities, enabling them to produce these types of artifacts with greater ease.''}    
% \end{quote}
% matt: this is not a participant quote; it's a quote from a paper (it is paper R8, referring to generation of writing artifacts like reports).

\subsubsection{Task Management \& Logistics} Within this stage, we aggregated all the secondary tasks that practitioners performed to ensure the smooth functioning of their roles. These tasks included planning, coordinating, and managing various resources and processes. Practitioners engaged with GenAI within this stage in two distinct ways. First, they used GenAI to streamline their overall workflow. This included using GenAI to help formulate clear project goals and develop concrete support mechanisms to achieve those goals~\cite{6}. %For example, a sci-fi writer from the same study shared how they leveraged GenAI to create a ``hero's journey'' document -- used by writers to track progress and plot elements. The writer then used ChatGPT to further break down the document into manageable steps, addressing various criteria such as deadlines and project budget.
Speech language pathologists used GenAI to optimize the documentation processes required at different stages of their workflow, such as recording intake forms and writing evaluation reports~\cite{7}. One speech language pathologist posted in an online community:

\begin{quote}
    \textit{``[...] I’m in the process of creating a Google Form for speech/language intake. After collecting responses (ensuring privacy by removing personal details), you can efficiently transfer this data into ChatGPT alongside a preset prompt template and your evaluation notes. [\dots] ''}
    %This approach can greatly expedite the report writing process, using AI to structure and incorporate information seamlessly into a Word document!''}
\end{quote}

Practitioners also used GenAI to delegate specific logistical activities within tasks to enhance work efficiency. One knowledge worker expressed in \cite{3}, ``\textit{It feels good to outsource this [kind of] work to ChatGPT because I don't enjoy it much}''~\cite{3}. Common delegated activities included summarizing material or simplifying dense content. For example, practitioners frequently used GenAI to condense their own written content for presentations or meeting notes while they focused on conveying critical business decisions. We found this strategy to be particularly prevalent in UX practitioners \cite{10} and knowledge workers \cite{3}. 
%In contrast, fact-checkers relied on GenAI tools to break down and restructure ``dense content'' into a format that would be simpler and more digestible for their end-users \cite{1}.

Another common logistical activity was delegating short but highly repetitive tasks to GenAI, such as generating repetitive content. Participants frequently offloaded such tasks to capable programs. For example, software developers used GenAI to generate boiler plate code~\cite{16}, while fact-checkers performed transcription and translation tasks using GenAI~\cite{1}. Similarly, many practitioners relied on GenAI to fix repetitive but predictable issues, such as a software developer using ChatGPT to write a bash script to change the capitalization convention of file names~\cite{11}. GenAI helped practitioners save significant time in these cases by reducing the need for context switching between different applications~\cite{8}.
%Streamlining or delegating work to GenAI allowed practitioners to focus on tasks they found more important or engaging, e.g., preparing for discussions or concentrating on more advanced aspects of their work~\cite{3}. 

\subsection{Transforming Role Attributes} \label{sec:role}
The second transformation involved practitioners using GenAI to enhance their roles, not only shifting their perspectives of these roles (cognitive crafting) but also expanding their role capacities (task crafting). Some of these transformations were temporary, wherein they augmented their abilities through GenAI for the purpose of completing a particular task. Other practices were more permanent, wherein practitioners to focus on professional development by learning skills relevant to their work. 

\subsubsection{Augmenting Abilities Through GenAI}
In many cases, practitioners used GenAI to expand the range of tasks they could perform without improving their own skills or acquiring additional knowledge in the long term. 
%We refer to this practice as augmenting abilities through GenAI and noticed a high prevalence in our sample studies.  

One area of work in which participants augmented their abilities with GenAI was in the realm of communication. An industrial designer in~\cite{18}, for example, described how CAD-generated images filled in visual details that allowed him to tell a story around the packaging he was designing. In another domain, fact-checkers described using GenAI to connect instantly with audience members on a broader scale, or to convert fact checks into an audio format to share through different modalities like TikTok~\cite{1}. Despite not knowing how to do these tasks before, practitioners could accomplish them successfully using GenAI. 

Other practitioners used GenAI to fill perceived gaps in their critical thinking abilities, such as by using GenAI to make external assessments (e.g., evaluating ideas for a startup)~\cite{21} or to improve their sense-making abilities. A designer in~\cite{18} used GenAI to reveal complex connections between their current work and previous work, connections that they were initially not aware of. A few practitioners used GenAI for more routine but core tasks where they experienced a disadvantage, such as proofreading manuscripts in a non-native language \cite{13}. GenAI also allowed practitioners to perform tasks beyond their skill set. 
%For example, fact-checkers used GenAI to generate search queries in languages they didn't speak or identify the sources of images~\cite{1}. 
This could take the form of GenAI filling in for missing expertise in constrained situations, e.g., ``\textit{In our startup, we don't have a dedicated UX writing role. Our designers often use ChatGPT to assess the appropriateness of the UX content in our design}''~\cite{2}.

\subsubsection{Learning Through GenAI}
Using GenAI also enabled practitioners to develop specific abilities and knowledge as a by-product of their dyadic interactions. Multiple developers argued that GenAI provided quicker and more effective access to information for knowledge acquisition than traditional search engines or reading the documentation directly~\cite{8,3}. 
%For example, a software developer claimed that,``\textit{what has been really helping me with this tool is precisely the process of trying to understand something that I’m not currently grasping about what a certain code is supposed to be doing}''~\cite{16}.  

Some practitioners exhibited explicit intent to use GenAI for learning purposes. Examples included developers using GenAI to learn how to do static type checking with type hints in Python~\cite{23}, exploring solutions to problems they'd never needed to solve before~\cite{4}, and understanding machine learning libraries~\cite{23}.

We also observed counterintuitive examples where, in attempts to improve productivity through GenAI, some practitioners skipped the learning process altogether. These practitioners experienced significant pressure to use the GenAI tools to decrease their project's delivery time but felt this compromised knowledge acquisition and skill development. For instance, a designer in~\cite{18} shared, ``\textit{there's no longer a day or a week of reading up for the project; instead I collect a bunch of materials and pretty much dump it into GPT-4.}''

The ability to use GenAI in this way reduce practitioners' sense of control and ownership around their work, sometimes leading to perceived roles evolving into assistants to GenAI. A graphic artist working with Midjourney found his role shifting to handing off ideas to the AI to illustrate, then fixing any errors, a change that he described as ``heartbreaking''~\cite{9}. At the same time, findings on practitioners' perceived roles with respect to GenAI as project managers~\cite{6} evoke the crafting practice of \textit{role re-framing}.

Others had more nuanced perspectives on these shifts, for example, perceiving they ways they manipulated GenAI outputs as a form of progressively retaking ownership~\cite{18}, or  
highlighting the value of uniquely human qualities in their work processes, e.g. empathy or artistic education~\cite{2,9}.

% Findings on humans' perceived roles, e.g., in \cite{6}, suggest the crafting practice of \textit{role re-framing}, shifting their sense of worth from their technical skills to their original ideas. As their experience grew, they began viewing themselves as managers of the GenAI tools. 

\subsection{Engaging in AI Managerial Labor}
\label{sec:aimanagerial}
The third form of transformation was the introduction of new planning and execution tasks that focused on maximizing the potential of GenAI in practitioners' work (task crafting) and integrating it fully into practitioners' workflows. We describe these tasks as \textit{AI managerial labor}.

One major form of AI managerial labor involved developing prompts, which required the additional work of establishing context. Tasks in this category included explaining big-picture goals, setting character limits for outputs, requesting specific output styles, providing examples of anticipated output, or providing a concrete starting point (e.g., code to start from), among others~\cite{4,6,16}. %like requesting a specific version of php, using descriptive variable names, etc. from 16
In rare cases, practitioners removed context by redacting sensitive information for compliance reasons from a prompt~\cite{18} or under-specified prompts to provoke surprising outputs~\cite{9}.
% One major form of AI managerial labor involved developing prompts. Practitioners frequently went to great lengths to furnish their prompts with context to help produce useful outputs for their work. For some, establishing context took the form of explaining their big-picture goals, setting character limits for outputs, requesting specific output styles (e.g., bulleted list), providing examples of anticipated output, or providing a concrete starting point (e.g., a code to start from)~\cite{4,6,16}. In rare cases, practitioners removed context, e.g., redacting sensitive information for compliance reasons from a prompt~\cite{18} or under-specifying prompts to provoke surprising outputs~\cite{9}. Some individuals were less willing to expend this kind of effort~\cite{15} and took an alternative route of curating ready-made prompts~\cite{14}. 
In a few instances, curation became a social endeavor of exchanging prompts; we discuss this in Section~\ref{sec:relational} on relational crafting. 

% We even identified certain practitioners who developed methodological practices around designing their prompts. For example, a developer in ~\cite{16} drafted their prompts in a separate tool before opening the GenAI, 

% \begin{quote}
%     \textit{``So there’s a step today that I often take before talking to ChatGPT, which is creating my prompt, creating my question. So, *I open a notepad*, think about what I’m going to put in the structure, and then I copy and paste it into the chat''}.
% \end{quote}

% A few practitioners viewed prompt designing as an iterative process. For instance, a software developer in~\citet{16} described receiving a wrong version of PHP code snippet from their initial prompt. The GenAI required additional clarification in follow-up prompts to get the syntax the developer was looking for. Others simply found it limiting to provide all the context in a single prompt, preferring to work incrementally towards their ultimate goal with smaller prompts~\cite{4}. Another developer observed,``\textit{Every time you prompt, you’re giving it clues to get closer to the idea you have in your head. The more words that you use, the closer the image can get to what you imagine~\cite{6}}.''  Interestingly, some practitioners included the GenAI in these conversations. For instance, some asked it about its capabilities and limitations or asked it to suggest prompts to use. Others even prompted it to critique its \textit{own} work~\cite{6}. \fixme{Underspecifying prompts [9]. Should it come here?}

A complementary form of AI managerial labor constituted refining GenAI outputs to fit in participants' workflows. Within this category, the most prominent activity consisted of verifying and correcting details in GenAI outputs. ``\textit{I don't trust the AI. \dots So, I have to read everything and validate what it's doing}''~\cite{15}. Beyond verification, some practitioners expended significant manual labor to utilize GenAI outputs. For instance, a UX designer from~\cite{2} spent time switching back and forth between GenAI tools and Photoshop to post-process images. In another domain,~\cite{17} described developers' post-processing pull requests authored by Copilot to either add missing information or remove superfluous context. Sometimes new processes emerged around this post-processing, such as maintaining a changelog to track changes made to the GenAI outputs. 

% In a more subtle form of AI managerial labor, practitioners sometimes exerted themselves in secondary ways around how they engaged with the AI. For example, one software developer described managing when the GenAI tool was on or off, because it gave proactive suggestions that could disrupt their workflow~{}.

AI managerial labor also involved configuring GenAI models at the system or application level, enabling practitioners to exert greater control over (a) data privacy, (b) model adaptation to their use case, and (c) the timing and context of the models' engagement with their workflows. One way practitioners took more control of the privacy of their data was by only using open-source models, as opposed to off-the-shelf models. They self-hosted them to prevent sensitive material from being exposed~\cite{1}. Practitioners also took control over adapting the model to their specific use cases by configuring settings. For example, a developer adjusted an API setting called the model temperature\footnote{https://platform.openai.com/docs/api-reference/chat/create\#chat-create-temperature}, which determines the randomness of the output given by the GenAI model, in order to generate more precise and helpful answers~\cite{6}.

%In a more involved example, a fact-checker fine-tuned GPT-3.5-Turbo with their own custom dataset to generate contextual queries. They could then leverage the contextual queries to perform more sophisticated queries in a different language that the practitioners did not speak ~\cite{1}. Fine-tuning like this is a process by which developers can provide specialized data to create their own custom version of a GenAI model which works better than the off-the-shelf model for their specific use case~\cite{ohm2024focusing}. 

While the above configurations were conducted at the systems-level, practitioners also spent time configuring GenAI tools at the application level, such as by managing when and how a GenAI provided them with suggestions in order to prevent disruptions to their work.  For instance, a developer found Copilot's proactive suggestion feature a source of interruption~\cite{16}. They kept Copilot off during the early stages of a project, saying, ``\textit{it tends to provide a lot of suggestions, which kind of hinders my thought process.}'' Note that this developer's experience contrasts with the affordances of standalone GenAI tools like ChatGPT that allow practitioners to choose when and for what activity to use them. 

% For instance, Copilot, unlike ChatGPT, was a more integrated interface in developers' code editor environments. Although this integration was useful in many ways, some developers described Copilot's proactive suggestion feature as interrupting their work. One developer, in \cite{16}, kept Copilot off during the early stages of a project, saying, 

% \begin{quote}
% \textit{``I believe that when I’m starting a project from scratch, I tend to keep it turned off because, as it still has very little context of what you’re doing, it tends to provide a lot of suggestions, which kind of hinders my thought process.''}
% \end{quote}

% A developer in \cite{22} shared how they selected integrated, proactive GenAI tools or standalone, reactive GenAI tools for different purposes:

% \begin{quote}
%     \textit{``I definitely use [Github Copilot’s Auto-Complete] much more [than Chat GPT]. I rarely actually go into Chat GPT... but I find that when I do use it, it’s more of like- it’s particular use case is kind of getting you unstuck from a spot rather than just integrating it into your regular flow \dots Sometimes it gives solutions and sometimes it’s just kind of to bounce ideas off.''}
% \end{quote}

% \input{4.0.FindingsTable}
\subsection{Transforming Professional Relationships}
\label{sec:relational}
We also saw preliminary evidence of practitioners using GenAI to shape their professional relationships (relational crafting). We believe it may be possible that the ability to interact with the GenAI tools in a two-way synchronous dialogue encouraged practitioners to implicitly have the same expectations from the GenAI that they might have from their human peers. Such expectations were reflected in practitioners' use of GenAI in specific aspects of their work that would usually be conducted through collaborations with peers. This involved participants asking GenAI to reflect on their works, to critique their work, and to provide alternative perspectives. 

For example, while working on their design project, a UI/UX designer used ChatGPT to set up checkpoints across their project life-cycle to help assess whether the current state of their artifact aligned with their envisioned outcome~\cite{6}. A more interesting use case was where fact-checkers in a Sudanese newsroom used GenAI to perform adversarial analysis by asking it to challenge the assumptions in articles written by the team~\cite{1}. Practitioners also established support networks on online platforms \cite{7,18}, such as Meta groups, to exchange specific prompts that enabled them to unlock new ways to engage with GenAI. 
%On several occasions, the comparisons between GenAI and human labor were made more explicit by practitioners anthropomorphizing the GenAI tools. These practitioners treated them as humans by acknowledging the human labor that they were doing with the tools, which was otherwise reserved for human roles.
Creative practitioners in~\cite{10} and~\cite{6} equated working with GenAI to ``\textit{quickly brainstorming with someone}.'' One practitioner shared:

\begin{quote}
    \textit{``I approached it similarly to how I would collaborate with someone. I could just go in with something half-baked and know that the system would ask me to clarify [if it needs it] \dots It really felt like a partnership [that] I found useful.''}
\end{quote}

GenAI could be used to widen the flexibility and perspectives of what might otherwise be possible through relational interactions with humans. For instance, a UX practitioner in \cite{21} fed empathy maps from an human interviewees to GenAI and asked it to emulate a persona based on them to support exploration across a wider range of scenarios. 
%This facilitated a more ``comprehensive understanding of the user’s perspective'' and ``enhanced the design exploration process''. 
In a more extreme case, the same practitioner experimented with assigning a non-human actor persona (e.g., technological systems or consumer objects) to GenAI to elicit design ideas. For example, the researcher asked ChatGPT to speak on behalf of a set of ESP32 microcontrollers, imagining how they (the microcontroller chips) might perceive their own role in the project~\cite{21}.

The perceived anthropomorphized contributions could directly impact practitioners' existing professional relationships. One way this materialized was when GenAI was used to replace ``\textit{taken for granted}'' tasks that were outsourced to other roles, such as transcription and translation \cite{1}. In more drastic situations, practitioners used GenAI to reduce their dependencies on different stakeholders. This was partially motivated by \textit{external} factors, such as intent to avoid wasting the time of specialized but overburdened roles or to save time and resources. For instance, UX designers used GenAI in their startup \cite{2} instead of hiring a UX writer to reduce the overhead costs. On the other hand, some of the reasons were \textit{internally} motivated, such as shortening the turn around time that was occurring because of practitioners' dependencies on other roles, such as experts. In \cite{9}, creative leaders, instead of waiting for artists' input, ``\textit{side-stepped}'' them and developed their own pitch decks for client communication using GenAI. 

Practitioners used GenAI to navigate personal and infrastructural constraints. For example, graphic designers in the Global South faced unreliable internet and power outages, making it difficult to brainstorm with clients~\cite{14}. To maintain pace of progress, practitioners compensated for the gap by brainstorming with ChatGPT. Practitioners also used GenAI to reduce interactions that contributed to increased stress in their jobs. For example, in \cite{5}, telephone operators working in the healthcare domain were tasked with the emotional labor of reaching out to older adults on a regular basis, having conversations with them, and recording any issues impacting their lives. When GenAI was introduced in this workflow it reduced the burden of operators doing these calls manually~\cite{5}.

% Some engaged in these processes to overcome personal constraints, such as infrastructural issues. For graphic designers in Global South, unreliable internet and power blackouts made it difficult for them to brainstorm with their clients and have multiple rounds of interactions to refine the project requirements. To work within the constraints, designers reduced their interactions with the client and instead conducted their brainstorming sessions with ChatGPT through prompts to achieve similar, if not perfect, goals. Practitioners also used GenAI to reduce those interactions that contributed to increased stress in their jobs. For example, in \cite{5}, telephone operators working in the healthcare domain were tasked with the emotional labor of reaching out to older adults on regular basis, having conversations with them, and recording any issues impacting their lives. GenAI was introduced in this workflow to reduce the burden of operators doing these calls manually. Interestingly, operators found GenAI extremely helpful in reducing their emotional labor and the resultant emotional stress that they were otherwise experiencing in their day-to-day work.  

\section{Discussion}
Our findings provide a nuanced view of how practitioners engaged in high-skill work have integrated GenAI to reshape their tasks, adjust their relationships, and ultimately shift their perceptions of their roles. These transformations also exposed practitioners to several underlying tensions, which we unpack here.
 
\subsection{The Dynamics of Task Transformation with GenAI}
Our insights show that practitioners primarily turned to GenAI to increase their overall work productivity and efficiency. However, in the process, they created a need for new forms of labor to manage the AI (see Section \ref{sec:aimanagerial}). In some ways, these findings reflect prior work around the long-term transformation of technology-driven automation on work \cite{autor2003skill}.  Autor\cite{autor2015there} argued that automation can increase the value of complementary tasks. 
%For example, the rise of AI-driven automation for data collection tasks has created demand for practitioners to make sense of this new data.
While this has traditionally led to a polarization of labor, where routine tasks are automated and create demand for tasks which require non-routine human abilities (abstract reasoning or manual interaction with messy environments), our findings suggest a shift with GenAI.

% At the same time, our findings diverge from the typical pattern described by \cite{autor2015there} in how this displacement occurs. \cite{autor2015there} suggested that the demand for routine well-structured tasks (also referred to as ``middle-skilled'' activities) tends to give way to tasks that complement automation and require unique human abilities, such as high-level abstract thinking or manual interaction with the messy, unpredictable world. 
% Our GenAI-specific findings demonstrate an opposite trend. 
In addition to routine tasks, some practitioners voluntarily gave GenAI abstract and creative tasks, while regularly taking on tasks to manage GenAI like fixing mistakes in its outputs. This departure reflects the unique capabilities of GenAI tools, which are more capable than previous forms of AI in completing tasks that require complex human abilities. Forms of AI managerial labor like verifying outputs still require human intuition. It remains to be seen whether these will also be susceptible to automation and how the market will value them.

% The pockets of tasks that were still exclusive to practitioners were those that involved human ability to infer contextual preferences and how those map to the outputs. This human quality characterizes much of the AI managerial labor that we observed.

\subsection{Arising Tensions in Shifting Role Identities}
Aforementioned transformations in tasks could also be accompanied by practitioners grappling with shifts in their perceived role identities.  Workers can perceive threats to their professional identity when they experience changes to their role~\cite{petriglieri2011under}. Workers often respond to such threats by either perceiving a sense of identity crisis or perceiving an opportunity for identity growth~\cite{zikic2016happens}. In the literature we surveyed, there were examples of both responses. At least one practitioner lamented the erosion of their role when they offloaded tasks that they perceived were central to their professional identity (see Section \ref{sec:role}). 
% This contributed to these practitioners perceiving GenAI to devalue their role and contributed to identity crises.

In contrast, others re-framed their roles or aspects of them more positively in response to GenAI, suggesting the possibility for identity growth through working with GenAI. These findings echo Zikic's \cite{zikic2016happens} work around threat perceptions of immigrant workers and their resultant experiences of identity crisis or growth when adapting to a new workplace. Future studies need to focus on designing support structures that help practitioners handle these threats in ways that lead to growth rather than identity crisis.

% -- role identity shift (start with periphery but leaking core tasks to genai)
% -- inability to control what tasks people give--more and more dependency
% -- tension with high vs low-level skills
% --capacity building vs augmentation
% --balance between increased productivity or freeing time to do other tasks vs increasing pressure/pace demand due to speed of GenAI

% role reduction happening in two ways: one by increasing their capability, they are now less dependent on someone else and also less dependent on themselves

% 2. human-machine collaboration

\subsection{The Versatility of GenAI \& Fractured Roles}
In order to redefine their tasks and roles, practitioners frequently restructured their work in innovative ways to engage effectively with GenAI. This happened across various stages of their workflows (see Section \ref{sec:workprocesses}) and through two primary approaches. First, practitioners segmented their own tasks into distinct parts, identifying elements that GenAI could independently accomplish, and then collaborated on these. Second, practitioners selectively undertook tasks typically managed by external stakeholders, completing them with GenAI’s support. This approach of compartmentalizing tasks with GenAI’s assistance began to resemble a form of piecework\cite{hagan1973piece}. Piecework is a process of decomposing complex work into smaller ``pieces'' that could be performed by individuals with varied skill sets~\cite{alkhatib2017examining}, with recent HCI applying this concept to modern crowd-work \cite{irani2015cultural}. 

It's possible that using GenAI as a pieceworker could alleviate some of the issues that typically arise around piecework. For example, GenAI's versatile features enabled practitioners to assign piecework to it while overseeing the tasks. Unlike traditional piecework dynamics, one might argue that situations where a practitioner chooses to parcel out elements of their work to GenAI gives them unprecedented agency in what aspects of their roles become piecework. In an ideal scenario, this could allow the practitioner to focus on more fulfilling aspects of their roles without exploiting other workers.

However, from practitioners' perspectives, transforming their tasks into piecework brought multiple uncertainties to their roles. One common uncertainty came from the fact that the affordances of the GenAI tool, rather than the preferences of the practitioner, tend to determine what aspects of work get off-loaded. If, for example, tweaking GenAI-generated images is significantly faster than a human artist producing those images, continuing to do this work manually can carry significant risk of falling behind. In addition, the potential for some roles to absorb pieces of other roles can introduce power asymmetries that undermine relationships between neighboring roles. Practitioners can also run the risk of transforming visible aspects of their work into invisible (e.g., prompt designing). This can lead to acrimonious relationships between workers and management~\cite{alkhatib2017examining}.

\subsection{Technology Crafting: A New Form of Job Crafting}

Finally, we observed several instances of job crafting with technology that do not fit the traditional job crafting framework. This form of crafting, which we label \textit{technology crafting}, is characterized by actions that are exclusively aimed at reconfiguring a specific \textit{technology} to improve one's work experience. In particular, this occurred through actions taken to manage or reconfigure GenAI tools that practitioners would not have otherwise engaged in except to integrate said tools into their practices. In these practices, GenAI became the focal point for modification and adoption, just like any other task or relationship.

Unlike prior research on AI \cite{Perez2022} in which users had little control over the structure and function of AI tools (e.g., algorithmic management through a tool), leaving employees to craft only reactively, GenAI was easily configurable and personalizable, supporting proactive crafting. At an extreme, this entailed tuning a model for a specific use case or setting up self-hosting to control sensitive data sharing. This kind of labor includes behaviors like curating prompts, tuning model parameters, or managing when to have a tool be active or inactive to better fit one's working style.

These forms of technology crafting differed from previously-studied forms in two key ways. First, it required continuous interaction with GenAI, imposing temporal demands. Second, the black-box nature of the technology introduced variability in control, making it challenging for professionals to accurately predict the outcomes.

\section{Limitations and Conclusion}
This work has a few limitations. To start, we focused our search on GenAI-enabled work practices performed in the HCI community. For this purspose, we limited ourselves to the ACM digital library. As more work emerges around how GenAI is being used, looking at broader research communities will help to tell a more comprehensive story. Further, the papers that we found relevant to our research objective were mostly qualitative. While this was appropriate to the nature of our question, quantitative survey studies can complement our narratives that we identified.

Finally, although GenAI tools are becoming accessible in fields beyond technology, the reviewed studies predominantly focused on technology-related occupations, highlighting a critical need for HCI studies to examine GenAI's impact across a broader range of professions.

In summary, this paper analyzed 23 papers to understand how GenAI is being used by practitioners to craft their jobs. We found that practitioners used GenAI to transform targeted aspects of the tasks they were performing, as well as to shape their roles and relationships. Based on our findings, we discussed how bottom-up usage of these tools was changing roles in unconventional ways, shifting task demand from high-level abstract thinking to more routine tasks, and facilitating the decomposition of roles into piecework. 
%We also suggest a need to expand the job crafting framework to consider ways in which practitioners craft the technology they use to transform their work experiences.

%
% ---- Bibliography ----
%
% BibTeX users should specify bibliography style 'splncs04'.
% References will then be sorted and formatted in the correct style.
%
\bibliographystyle{splncs04}
\bibliography{mainbib}

\end{document}